\documentclass[conference]{IEEEtran}
\IEEEoverridecommandlockouts
\usepackage{cite}
\usepackage{amsmath,amssymb,amsfonts}
\usepackage{algorithmic}
\usepackage{graphicx}
\usepackage{textcomp}
\usepackage{xcolor}
\usepackage{hyperref}
\usepackage{breakurl}

\def\BibTeX{{\rm B\kern-.05em{\sc i\kern-.025em b}\kern-.08em
    T\kern-.1667em\lower.7ex\hbox{E}\kern-.125emX}}

\makeatletter
\newcommand{\linebreakand}{%
  \end{@IEEEauthorhalign}
  \hfill\mbox{}\par
  \mbox{}\hfill\begin{@IEEEauthorhalign}
}
\makeatother

\begin{document}

%\title{Analyzing Cryptocurrencies and Identifying Risky Tokens using Machine Learning and On-chain Parameters
%\title{Cryptocurrency Analysis and AI-assisted Identification of Risky Tokens
\title{AI-Assisted Investigation of On-Chain Parameters: Risky Cryptocurrencies and Price Factors
}

\author{\IEEEauthorblockN{Abdulrezzak Zekiye}
\IEEEauthorblockA{\textit{Department of Computer Engineering} \\
\textit{Koç University}\\
Istanbul, Türkiye \\
azakieh22@ku.edu.tr}
\and
\IEEEauthorblockN{Semih Utku}
\IEEEauthorblockA{\textit{Department of Computer Engineering} \\
\textit{Dokuz Eylül University}\\
Izmir, Türkiye \\
semih@cs.deu.edu.tr}
\and
\linebreakand
\IEEEauthorblockN{Fadi Amroush}
\IEEEauthorblockA{\textit{Chief Technology Officer} \\
\textit{Niuversity}\\
Berlin, Germany \\
fadi.amr@niuversity.com}
\and
\IEEEauthorblockN{Öznur Özkasap}
\IEEEauthorblockA{\textit{Department of Computer Engineering} \\
\textit{Koç University}\\
Istanbul, Türkiye \\
oozkasap@ku.edu.tr}
}

\maketitle

\begin{abstract}
Cryptocurrencies have become a popular and widely researched topic of interest in recent years for investors and scholars. In order to make informed investment decisions, it is essential to comprehend the factors that impact cryptocurrency prices and to identify risky cryptocurrencies. This paper focuses on analyzing historical data and using artificial intelligence algorithms on on-chain parameters to identify the factors affecting a cryptocurrency's price and to find risky cryptocurrencies. We conducted an analysis of historical cryptocurrencies' on-chain data and measured the correlation between the price and other parameters. In addition, we used clustering and classification in order to get a better understanding of a cryptocurrency and classify it as risky or not. The analysis revealed that a significant proportion of cryptocurrencies (39\%) disappeared from the market, while only a small fraction (10\%) survived for more than 1000 days. Our analysis revealed a significant negative correlation between cryptocurrency price and maximum and total supply, as well as a \textcolor{black}{weak} positive correlation between price and 24-hour trading volume. Moreover, we clustered cryptocurrencies into five distinct groups using their on-chain parameters, which provides investors with a more comprehensive understanding of a cryptocurrency when compared to those clustered with it. Finally, by implementing multiple classifiers to predict whether a cryptocurrency is risky or not, we obtained the best f1-score of 76\% using K-Nearest Neighbor.
\end{abstract}


\section{Introduction}
Cryptocurrencies have gained significant attention from investors in recent years, with a market capitalization of approximately 1,190 billion USD and a trading volume of about 44 billion USD, according to CoinMarketCap \cite{coinmarketcap}. However, it is crucial to understand the characteristics of cryptocurrencies to make informed investment decisions. Unlike traditional currencies, cryptocurrencies are decentralized and not controlled by governments or banks. The price of a cryptocurrency is influenced by off-chain parameters such as news, social media rumors, regulations, and the reputation of the team behind the cryptocurrency, which are unpredictable and cannot be relied upon to make stable investment decisions. In contrast, on-chain parameters, which are saved in the cryptocurrency's blockchain, are more stable and predictable and play a crucial role in affecting the cryptocurrency's price. In \cite{zekiyecrypto}, it was concluded that the top 10 important on-chain parameters for cryptocurrencies are circulating supply, market cap, volume over 24 hours, percent of total supply circulating, total staked, staking reward, whales percentage, total value locked, number of market pairs, and date added. Circulating Supply refers to the coins available in the market for public use. \textcolor{black}{Market capitalization, commonly known as market cap, is the total value of the publicly available coins to trade with. We can calculate market cap by multiplying the price of the coin by its circulating supply (Equation.\eqref{eq.marketcap}). Volume over 24 hours is the value of transactions that occurred in the past 24 hours. Percent of Total Supply Circulating (PTSC) indicates the percentage of coins available for public use out of the total existing coins of a particular cryptocurrency and it is calculated using Equation.\eqref{eq.percentage} \cite{zekiyecrypto}}. Total Staked is the total value staked as a percentage of circulating supply, while Staking Reward is the reward received by stakers as a percentage of their staked asset. Whales' Percentage represents the percentage of circulating supply held by whales, where a whale is an address holding at least 1\% of the circulating supply \textcolor{black}{per the definition of IntoTheBlock \cite{intotheblock}}. Total Value Locked denotes the value of cryptocurrencies locked to take on loans. The Number of Market Pairs indicates the number of pairs that can be exchanged with cryptocurrencies. Finally, the Date Added refers to the date the cryptocurrency was added to the market. It is important to note that the number of market pairs is an off-chain parameter but was included in the study due to its importance in giving trust to investors \cite{zekiyecrypto}.

\begin{equation}
Market\ cap = price * circulating\ supply
\label{eq.marketcap}
\end{equation}

\begin{equation}
PTSC = \frac{Circulating\ Supply}{Total\ Supply}
\label{eq.percentage}
\end{equation}

Several studies have been conducted to develop recommendation systems and predict the prices of cryptocurrencies. In \cite{chiang2021development}, the researchers proposed a cryptocurrency investment decision support system that utilizes a Recurrent Neural Network (RNN) and a Long Short-Term Memory (LSTM) to predict future prices based on daily cryptocurrency prices. In \cite{zekiyecrypto}, a decision support system was developed to help investors choose suitable cryptocurrencies based on their preferences using on-chain features and the Analytic Hierarchy Process algorithm. Other studies focused on detecting risks and manipulation in the cryptocurrency market. \textcolor{black}{Researchers in \cite{enoksen2020understanding} tried to understand the state of a cryptocurrency where the price increases dramatically beyond its real value causing what is called a bubble. The study revealed a positive correlation between the presence of bubbles and increased volatility, trading volume, and transaction levels}. In \cite{li2021cryptocurrency}, the involvement of gambling and overconfident investors in pump-and-dump schemes, and their detrimental effects on liquidity and prices were highlighted. Researchers in \cite{sovbetov2018factors} \textcolor{black}{ studied the factors that affect the price of five cryptocurrencies: Bitcoin, Ethereum, Dash, Litcoin, and Monero. They found} that cryptomarket-related factors such as market beta, trading volume, and volatility are significant determinants of the prices of all five cryptocurrencies in the short- and long-run, while the attractiveness of cryptocurrencies also matters for their price determination, but only in the long-run. \textcolor{black}{In \cite{raza2021digital}, researchers used deep learning to forecast bitcoin prices and mentioned that such forecasting can be used to mitigate business risks. In their model, only price data were used to forecast future price value.}

It is noteworthy that there has been a lack of efforts to detect manipulation and identify risky cryptocurrencies using their on-chain parameters. \textcolor{black}{Identifying a cryptocurrency as risky in our context means that there is a high risk in investing in it. More precisely, we labeled the cryptocurrencies that disappeared from the market as risky ones and trained our model based on them. As a result, a risky cryptocurrency means that it might disappear from the market in the future}. 

In this paper, we aim to identify the parameters utilized for manipulating cryptocurrency prices by analyzing the correlation between the on-chain parameters and price. We aim also to apply artificial intelligence techniques to classify cryptocurrencies into risky or non-risky categories, where risky means that such a cryptocurrency might disappear in the future. Moreover, clustering methods are employed to enhance the comprehension of a cryptocurrency's risk level, particularly when it is clustered with a known risky cryptocurrency.

The contributions of this paper are as follows:

\begin{itemize}
   \item Identifying parameters used for manipulating cryptocurrency prices through analysis of historical on-chain data and measuring the correlations between price and other parameters. \textcolor{black}{By identifying those parameters, investors can determine whether a cryptocurrency can be easily manipulated or not and thus make better investment decisions}. From our analysis of the data spanning from 2013 to January 1st, 2022, it was discovered that a significant portion of cryptocurrencies (39\%) have exited the market, with only a small fraction (10\%) having lasted for more than 1000 days. Furthermore, by studying the correlation between the price and the other on-chain parameters for the cryptocurrencies data \textcolor{black}{between January 1st, 2020 and January 1st, 2022}, we observed a strong negative correlation between cryptocurrency price and maximum and total supply, and a moderate positive correlation between price and 24-hour trading volume. 
   
   \item Applying artificial intelligence techniques to classify cryptocurrencies into risky or non-risky categories based on on-chain parameters. Seven classifiers were utilized to classify whether a cryptocurrency is risky or not, \textcolor{black}{where risky means that it might exit the market in the future}. The K-Nearest Neighbor classifier demonstrated the best f1-Score of approximately 76\%, indicating its effectiveness in identifying risky cryptocurrencies.

    \item Utilizing clustering methods on on-chain parameters to improve understanding of a cryptocurrency's risk level, particularly when clustered with a known risky cryptocurrency. We utilized K-means algorithm and the Elbow method to cluster cryptocurrencies based on their on-chain features. By using the Within-Cluster Sum of Square (WCSS) value, the optimal number of clusters was determined to be 5. \textcolor{black}{It is important to notice that the number of clusters does not represent risk levels but represents different groups of cryptocurrencies}.
\end{itemize}

\section{Analyzing Historical Data}

The data utilized in this study was collected from CoinMarketCap's history API, covering the period from 2013 to January 1st, 2022. However, it should be noted that the historical data only includes information on volume 24 hours, market cap, maximum supply, total supply, circulating supply, and price. The number of market pairs data is only available from May 10th, 2019, onwards.

\begin{figure}[t]
\centerline{\includegraphics[width=\linewidth]{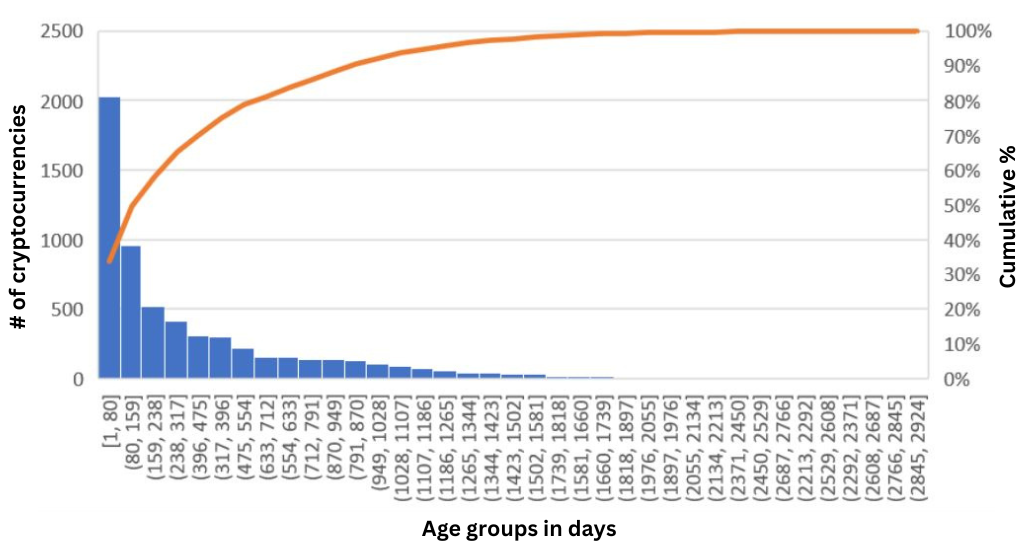}}
\caption{Pareto Chart Illustrating Disappeared Cryptocurrencies as of January 1st, 2022.}
\label{fig.pareto_disappeared}
\end{figure}

After examining historical data, we observed that certain cryptocurrencies have disappeared from the market. "Disappeared" refers to a cryptocurrency that is no longer available for trading, and in some instances, has undergone a name or symbol change. Out of the 15,349 unique name-symbol combinations, only 9,310 remained, indicating that roughly 39.34\% of cryptocurrencies have disappeared. Fig. \ref{fig.pareto_disappeared} displays a Pareto chart of the vanished cryptocurrencies between 2013 and January 1st, 2022, illustrating their lifetimes in descending order of frequency, with a cumulative line on a secondary axis as a percentage of the total. A cryptocurrency's lifetime was calculated by measuring the difference between the first day it entered the market and the last day it was available. We observed that 40\% of disappeared cryptocurrencies had a lifetime of fewer than 80 days, while approximately 75\% had a lifespan of less than a year.

\begin{figure}[t]
\centerline{\includegraphics[width=\linewidth]{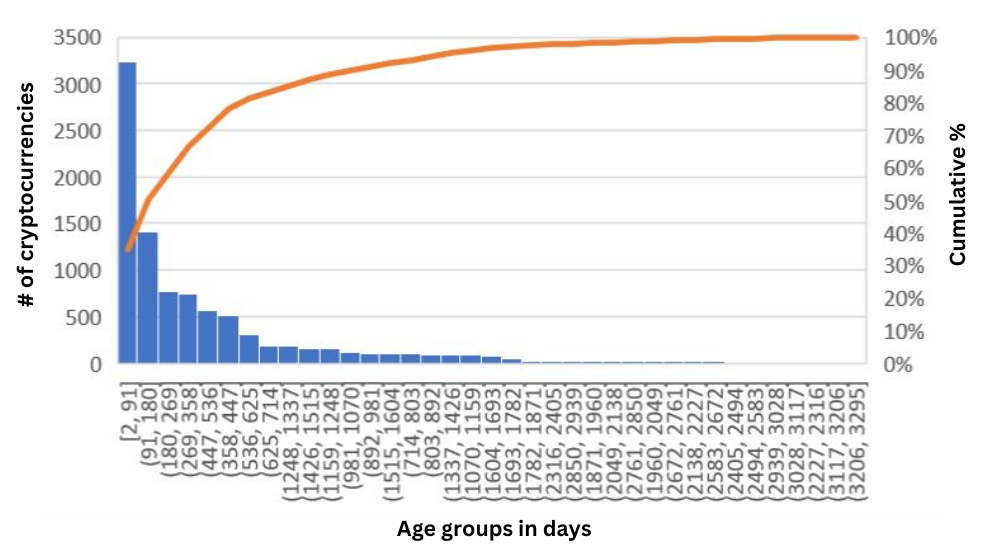}}
\caption{Pareto Chart Illustrating existing Cryptocurrencies till January 1st, 2022.}
\label{fig.pareto_existing}
\end{figure}

Displayed in Fig. \ref{fig.pareto_existing} are the equivalent calculations but for cryptocurrencies still present in the market as of February 1st, 2022. It is noteworthy that around 10\% of these cryptocurrencies have been available for trading for more than 1000 days.

In order to measure the significance of parameters and their impact on cryptocurrency pricing, we calculated correlations between price and Max Supply, Total Supply, Circulating Supply, Volume over 24 hours, and Percentage of Total Supply Circulating using data from January 1st, 2020 to January 1st, 2022. We employed Pearson \cite{pearson1920notes}, Kendall Tau \cite{kendall1938new}, and Spearman \cite{spearman1961proof} correlations to measure these relationships \textcolor{black}{while considering the interpretation of the correlation coefficient in Table \ref{tab.correlation_interpretation} \cite{napitupulu2018analysis}}. While Pearson examines linear relationships between parameters, Kendall Tau and Spearman do not require such a relationship. Table \ref{tab.correlations} shows the Pearson, Kendall Tau, and Spearman correlations between price and the aforementioned parameters. Our observations reveal no linear (Pearson) correlation between price and the other parameters. However, Kendall Tau correlation indicates a \textcolor{black}{medium} negative correlation between maximum supply and total supply and price, as well as a weak positive correlation between volume and price. Furthermore, according to Spearman correlation, we observe:
\begin{enumerate}
    \item Strong negative relationship between maximum supply and price.
    \item Strong negative relationship between total supply and price.
    \item \textcolor{black}{Medium} positive relationship between volume over 24 hours and price.
\end{enumerate}

\begin{table}[htbp]
\caption{Interpretation of Correlation Coefficient \cite{napitupulu2018analysis}}
\begin{center}
\begin{tabular}{|c|c|}
\hline
\textbf{Coefficient Interval}&\textbf{Correlation Interpretation}\\
\hline
0.00-0.199 &  Very Weak \\
\hline
0.20-0.399 &  Weak \\
\hline
0.40-0.599 &  Medium \\
\hline
0.60-0.799 &  Strong \\
\hline
0.80-1.000 &  Very Strong \\
\hline  
\end{tabular}
\label{tab.correlation_interpretation}
\end{center}
\end{table}

\begin{table*}[htbp]
\caption{Correlations between price and on-chain parameters Using Pearson, Kendall Tau, and Spearman Methods}
\begin{center}
\begin{tabular}{|c|c|c|c|c|c|}
\hline
\textbf{}&\textbf{Maximum Supply} &\textbf{Total Supply}&\textbf{Circulating Supply}&\textbf{Volume over 24 hours}&\textbf{Percentage of Total Supply Circulating}\\
\hline
\textbf{Pearson} & 0.0014 &  -0.00137 &   -0.00042 &  0.00362 &  0.00252 \\
\hline
\textbf{Kendall Tau} &  -0.47424 & -0.46843 & -0.01480 & 0.27754 & 0.09169 \\
\hline
\textbf{Spearman} & -0.63634 & -0.63254 & 0.00376 & 0.40029 & 0.12903 \\
\hline
\end{tabular}
\label{tab.correlations}
\end{center}
\end{table*}

We conducted correlation experiments between the mean and standard deviation of price and the mean and standard deviation of maximum supply, total supply, volume over 24 hours, and percentage of total supply circulating. Table \ref{tab.mean_correlations_stat} displays the correlation results between the mean of price and other parameters, while Table \ref{tab.dev_correlations_stat} shows the correlation results between the standard deviation of price and other parameters.

We can interpret the standard deviation as the level of fluctuation. For instance, the standard deviation of price reflects the extent of price fluctuation. It is noticeable that Spearman correlation demonstrates the highest values. When looking at the Spearman correlation between the mean price and other parameters, there is a \textcolor{black}{strong} negative correlation with the mean of maximum supply and the mean of total supply. The Spearman correlation between the mean price and the mean of volume over 24 hours and the standard deviation of volume over 24 hours is of a \textcolor{black}{weak} but nearly \textcolor{black}{medium} value. A similar level of Spearman correlation is present between the standard deviation of price and the aforementioned parameters.

\textcolor{black}{Finally, we can see the Spearsman correlations between Maximum Supply, Total Supply, Circulating Supply, Volume over 24 hours, Market Cap, and Number of Market Pairs in Table \ref{tab.other_corr}. A noticeable very strong correlation between market cap and circulating supply exists and it is expected since circulating supply is used to calculate the market cap. In addition, we can see a strong correlation between volume over 24 hours and the number of market pairs and market cap.}

\begin{table*}[htbp]
\caption{Correlation between the mean of price and other statistical parameters}
\begin{center}
\begin{tabular}{|c|c|c|c|c|c|c|c|c|}
\hline
\textbf{}& \parbox[c]{1.5cm}{\textbf{Mean of maximum supply}} &\parbox[c]{1.5cm}{\textbf{The standard deviation of maximum supply}}&\parbox[c]{1.5cm}{\textbf{Mean of
the total
supply
}}&\parbox[c]{1.5cm}{\textbf{The standard
deviation of
the total
supply}}&\parbox[c]{1.5cm}{\textbf{Mean of
volume over
24 hours
}} & \parbox[c]{1.5cm}{\textbf{The
standard
deviation of
volume over
24 hours
}} & \parbox[c]{1.5cm}{\textbf{Mean of the
percentage of
total supply
circulating}} & \parbox[c]{1.7cm}{\textbf{The standard
deviation of the
percentage of total
supply circulating
}}\\
\hline
\textbf{Pearson} & -0.00428 & -0.00107 & -0.00436 & -0.00106 & 0.09352 & 0.00371 & 0.00601 & -0.00221 \\
\hline
\textbf{Kendall Tau} & -0.54445 & 0.08691 & -0.55059 & 0.10719 & 0.23039 & 0.20466 & 0.20744 & 0.22959 \\
\hline
\textbf{Spearman} & -0.70568 & 0.11123 & -0.71536 & 0.14280 & 0.34104 & 0.30617 & 0.27806 & 0.30110 \\
\hline
\end{tabular}
\label{tab.mean_correlations_stat}
\end{center}
\end{table*}

\begin{table*}[htbp]
\caption{Correlation between the standard deviation of price and other statistical parameters}
\begin{center}
\begin{tabular}{|c|c|c|c|c|c|c|c|c|}
\hline
\textbf{}& \parbox[c]{1.5cm}{\textbf{Mean of maximum supply}} &\parbox[c]{1.5cm}{\textbf{The standard deviation of maximum supply}}&\parbox[c]{1.5cm}{\textbf{Mean of
the total
supply
}}&\parbox[c]{1.5cm}{\textbf{The standard
deviation of
the total
supply}}&\parbox[c]{1.5cm}{\textbf{Mean of
volume over
24 hours
}} & \parbox[c]{1.5cm}{\textbf{The
standard
deviation of
volume over
24 hours
}} & \parbox[c]{1.5cm}{\textbf{Mean of the
percentage of
total supply
circulating}} & \parbox[c]{1.7cm}{\textbf{The standard
deviation of the
percentage of total
supply circulating
}}\\
\hline
\textbf{Pearson} & -0.00213 &  -0.00014 &  -0.00215 &  -0.00015 & 0.00902 &  0.00014 &  -0.00324 &  -0.00130 \\
\hline
\textbf{Kendall Tau} & -0.53146 & 0.12428 & -0.53893 & 0.13971 & 0.23198 & 0.21908 & 0.22161 & 0.25433 \\
\hline
\textbf{Spearman} & -0.69011 & 0.15828 & -0.70099 & 0.18461 & 0.34413 & 0.32804 & 0.29731 & 0.33242 \\
\hline
\end{tabular}
\label{tab.dev_correlations_stat}
\end{center}
\end{table*}

\begin{table*}[!htbp]
    \caption{Spearsman correlations between Maximum Supply, Total Supply, Circulating Supply, Volume over 24 hours, Market Cap, and Number of Market Pairs}

    \centering
    \begin{tabular}{|c|c|c|c|c|c|c|}
    \hline
          & \textbf{Maximum Supply} & \textbf{Total Supply} & \textbf{Circulating Supply} & \parbox[c]{1.8cm}{\textbf{Volume over 24 hours}} & \textbf{Market Cap} & \parbox[c]{1.8cm}{\textbf{Number of Market Pairs}} \\ \hline
        \textbf{Maximum Supply} & 1 & 0.39174 & 0.04483 & 0.08657 & -0.03969 & 0.00434 \\ \hline
        \textbf{Total Supply} & 0.39174 & 1 & 0.51003 & 0.14479 & 0.41273 & 0.26707\\ \hline
        \textbf{Circulating Supply} & 0.04483 & 0.51003 & 1 & 0.20325 & 0.92143 & 0.42596 \\ \hline
        \textbf{Volume over 24 hours} & 0.08657 & 0.14479 & 0.20325 & 1 & 0.34674 & 0.62065 \\ \hline
        \textbf{Market Cap} & -0.03969 & 0.41273 & 0.92143 & 0.34674 & 1 & 0.52127 \\ \hline
        \textbf{Number of Market Pairs} & 0.00433 & 0.26707 & 0.42596 & 0.62065 & 0.52127 & 1 \\ \hline
    \end{tabular}
    \label{tab.other_corr}

\end{table*}

\section{Clustering Cryptocurrencies}
In \cite{zekiyecrypto}, a cryptocurrency investment decision support system using on-chain parameters was proposed and implemented. However, some investors may have difficulty selecting the appropriate parameters and answering the questions accurately in the proposed system. In such cases, clustering can be a solution. Furthermore, clustering enables determining the category in which a cryptocurrency falls, allowing investors to exercise caution if a cryptocurrency is clustered with a known risky one.

\begin{figure}[t]
\centerline{\includegraphics[width=\columnwidth]{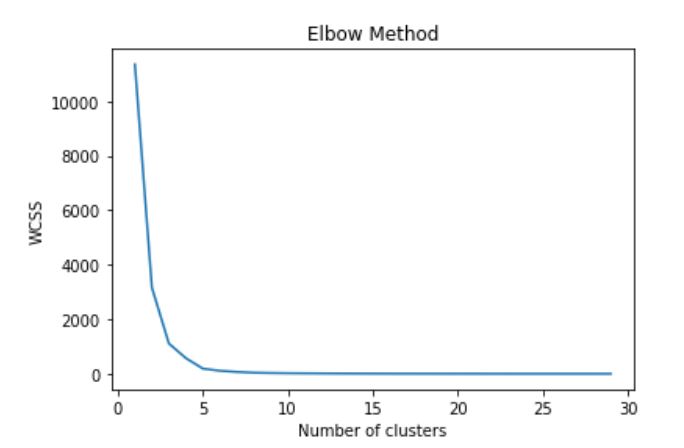}}
\caption{Within-Cluster Sum of Square (WCSS) values for the studied cryptocurrencies with varying number of clusters between 1 and 30}
\label{fig.elbow}
\end{figure}

Clustering involves grouping data with similar features into clusters. In this study, we used the K-means algorithm \cite{macqueen1967classification} to cluster the cryptocurrencies based on their on-chain features. The K-means algorithm uses the average squared distance between data points to form clusters. To initialize the algorithm, we used the k-means++ \cite{arthur2007k} to select cluster centers and speed up convergence. We normalized the features by dividing each one by its maximum value and then clustered the cryptocurrencies into five clusters using the PHP-ML library.
To determine the optimal number of clusters, we employed the Elbow method \cite{syakur2018integration}. We varied the number of clusters from 1 to 30 and calculated the Within-Cluster Sum of Square (WCSS) value. WCSS (Within-Cluster Sum of Squares) measures the sum of the squared distances between each data point within a cluster and its centroid. As shown in Fig. \ref{fig.elbow}, the WCSS value stopped changing significantly at the elbow point, which occurred when the number of clusters was 5. Therefore, 5 clusters were deemed the optimum number for our system. \textcolor{black}{We applied the clustering method based on daily data. In our experiment, we used the cryptocurrencies data on January 1st, 2022 with the following on-chain parameters: Market Cap, Volume over 24 hours, Number of Market Pairs, Percent of Total Supply Circulating, Total Value Locked, Staking Reward, Total Staking Percentage, and Whales Percentage.  Table \ref{tab.clusters} shows the clusters we obtained on the mentioned data where we included only the cryptocurrencies that have no missing values in the used on-chain parameters.}   

\begin{table}[htbp]
\caption{Cryptocurrency clustering on January 1st, 2022 using Market Cap, Volume 24h, Num Market Pairs, Percent of Total Supply Circulating, Total Value Locked, Staking Reward, Total Staking Percentage, and Whales Percentage. Cryptocurrencies with missing values were excluded.
}
\begin{center}
\begin{tabular}{|c|c|}
\hline
\textbf{Cluster \#} & \textbf{Cryptocurrencies} \\
\hline
\textbf{1} & Ethereum\\
\hline

\textbf{2} & Wabi\\
\hline

\textbf{3} & Olympus v2\\
\hline
\textbf{4} & \parbox[c]{6cm}{Polygon, Axie Infinity, Aave, The Graph, yearn.finance, IoTeX, SushiSwap, 0x, SwissBorg, Ontology, Cartesi, Dusk Network, Origin Protocol, Illuvium, Orbs, API3, IDEX, Akropolis, NULS, Bytom, Fusion, Dock, Remme, Bela, Livepeer, DODO, Fantom, Sentinel, Neutrino USD, Atomic Wallet Coin, Band Protocol, Curve DAO Token}\\
\hline

\textbf{5} & \parbox[c]{6cm}{BNB, Solana, Cardano, Polkadot, Avalanche, Cosmos, Internet Computer, Elrond, Tezos, Flow, Kusama, Oasis Network, Waves, Mina, Secret, Decred, WAX, Kava, Synthetix, SKALE Network, Hive, Phantasma, Persistence, Akash Network, Divi, Ark, Telos, PEAKDEFI, Switcheo, e-Money, Hydra, Particl, Peercoin, ChainX, Callisto Network, Tachyon Protocol, Minter Network, Blocknet, OKCash, Veil, Datamine, Rapids, Tendies, Savix, FireStarter, Terra, Lisk, Wagerr, Unification, Stake DAO, Starname, NEAR Protocol, IRISnet, CertiK, v.systems, HTMLCOIN, DeFiChain, Mirror Protocol, LTO Network, Stafi, Harmony, PIVX, Stacks, Zilliqa, Beefy Finance, Nexus, Trittium, Bitcoin Green, Algorand, PancakeSwap, Ardor, Edgeware, Phore, TokenPay, Pinkcoin, Crypto.com Coin, TRON, THORChain, IOST, TomoChain, Aion, FLETA, ReddCoin, Nxt, BlackCoin, CloakCoin, EOS, Celo, NEM, 1inch Network, Qtum, ICON, COTI, HEX, Kyber Network Crystal v2, STAKE, Energi, MANTRA DAO, InsurAce, Validity, Navcoin, Neblio, Enecuum, ChangeNOW Token, SmartCash, Wanchain, Thorstarter, Kalamint} \\
\hline
\end{tabular}
\label{tab.clusters}
\end{center}
\end{table}

\section{Predicting Risky Cryptocurrencies}

We observed that certain cryptocurrencies share either a name or symbol. To prevent any potential confusion, we combined each cryptocurrency's name and symbol with an underscore. The libraries we employed - Keras \cite{chollet2015keras}, NumPy \cite{harris2020array}, Matplot \cite{Hunter:2007}, and scikit-learn \cite{scikit-learn} - do not allow for null values. Consequently, we substituted null values with the mean value of their corresponding column. However, for the max supply values, null indicates infinity, which cannot be processed during classification. Filling these null values with the mean value would be incorrect. To address this issue, we replaced null values within the maximum max supply multiplied by one thousand. Ultimately, we normalized the values utilizing the mean normalizer, which is defined by Equation.\eqref{eq.normalizing}.
\begin{equation}
Normalized\ value = \frac{value - mean\ value}{standard\ deviation}
\label{eq.normalizing}
\end{equation}
The data used for training was divided into 80\% for training and 20\% for testing. \textcolor{black}{The experiments were conducted using Google Colab. However, due to a constraint of limited resources, specifically 12 GB of RAM, we were unable to process the entire dataset. Consequently, we had to limit the data processing to historical data spanning from 2013 to 2020}. Random splitting was employed for all classifiers, except for the Long-Short Term Memory classifier.

We employed seven different classifiers to determine whether a cryptocurrency is risky or not. In our experiments, we defined a cryptocurrency that has disappeared as risky, and the one that is still in the market as not risky. We did not say exactly whether a cryptocurrency is going to disappear or not as it is impossible to predict logically due to various parameters affecting such a prediction. The input for the classifiers included the normalized and cleaned historical data of the following parameters: Price, Maximum Supply, Total Supply, Circulating Supply, Volume over 24 hours, Market Cap, and Percentage of Total Supply Circulating. We used four metrics to evaluate the classifiers: precision, recall, accuracy, and f1-score, as defined by Equations.\eqref{eq.precision}, \eqref{eq.recall}, \eqref{eq.accuracy}, and \eqref{eq.f1}, respectively, \textcolor{black}{where TP is true positive, TN is true negative, FP is false positive, and FN is false negative \cite{dalianis2018evaluation}}. Among the four metrics, we considered f1-score to choose the best classifier since the classes (risky and not risky) are imbalanced.
\begin{equation}
Precision = \frac{TP}{TP + FP}
\label{eq.precision}
\end{equation}

\begin{equation}
Recall = \frac{TP}{TP + FN}
\label{eq.recall}
\end{equation}

\begin{equation}
Accuracy = \frac{TP + TN}{TP + FP + TN + FN}
\label{eq.accuracy}
\end{equation}

\begin{equation}
F_1\ Score = 2 \times \frac{Precisio\cdot Recall}{Precision+Recall}
\label{eq.f1}
\end{equation}

\begin{figure}[htbp]
\centerline{\includegraphics[width=\columnwidth]{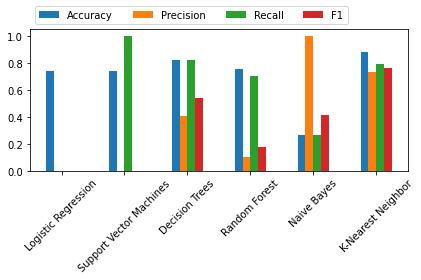}}
\caption{Classification results of Logistic Regression, Support Vector Machines, Decision Trees, Random Forests, Naïve Bayes, and K-Nearest Neighbor classifiers on the test dataset.
}
\label{fig.classification_results}
\end{figure}

Various classifiers with their default settings from the scikit-learn library were employed in this study. The classifiers used include Logistic Regression \cite{cramer2002origins}, Support Vector Machines \cite{gonen2008multiclass}, Decision Trees \cite{myles2004introduction}, Random Forests \cite{breiman2001random}, Naïve Bayes \cite{webb2010naive}, and K-Nearest Neighbor \cite{peterson2009k}. The performance outcomes of these classifiers are presented in Fig. \ref{fig.classification_results}. \textcolor{black}{We notice from the figure that precision, recall, and f1-score are zeros for the logistic regression classifier. In addition, precision and f1-score are zeros for the support vector machine classifier}.

\begin{figure*}[htbp]
\centerline{\includegraphics[width=\linewidth]{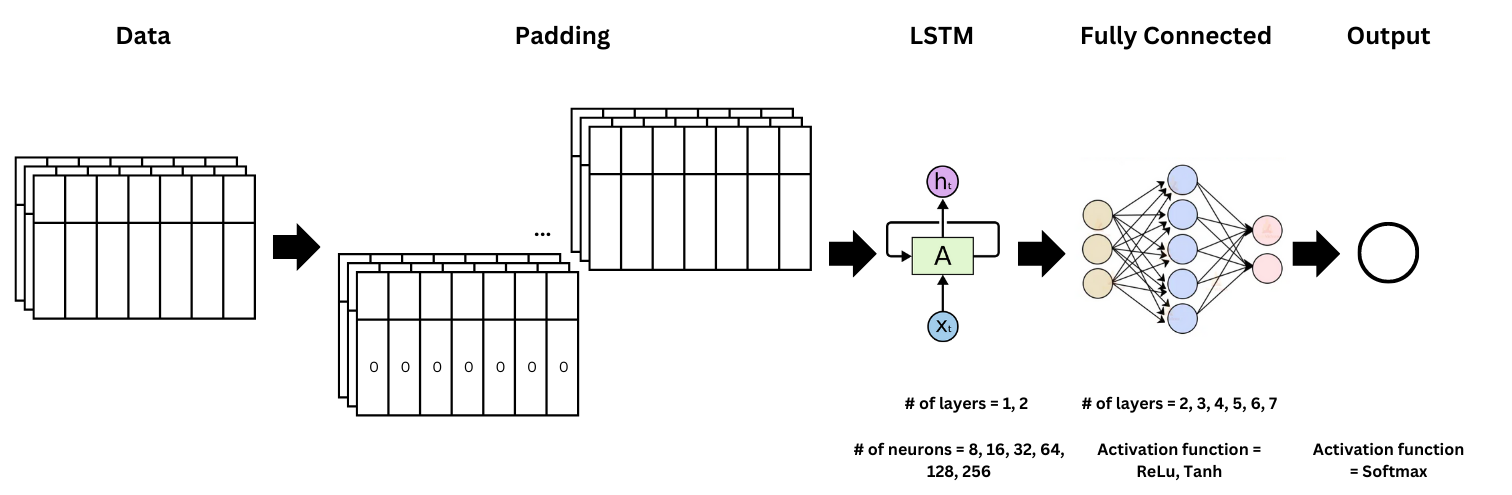}}
\caption{The LSTM architecture developed for cryptocurrency classification.
}
\label{fig.lstm}
\end{figure*}

We utilized Long-Short Term Memory (LSTM), \cite{yu2019review}, to classify cryptocurrencies based on unique symbol-name pairs, forming a time stream. The longest time stream, belonging to Bitcoin, spanned 2801 days. To account for LSTM's inability to process variable time series, we added padding to each coin's series with 7 features of leading zeros. Various LSTM architectures were tested, all yielding similar metrics. Fig. \ref{fig.lstm} displays the LSTM architecture used, and Table \ref{tab.lstm} provides the corresponding model metrics. We experimented with 1 and 2 LSTM layers, each with 8, 16, 32, 64, 128, and 256 units. Hidden layers were also tested, ranging from 2 to 7 layers, with the number of neurons starting at the same number as the LSTM units and halving for each subsequent layer. The output was a single neuron with Softmax activation function for binary classification, with 0 indicating a non-risky cryptocurrency and 1 indicating a risky one. Tanh and ReLu \cite{ramachandran2017searching} activation functions were both tested in the hidden layers, yielding similar results.

\begin{table}[htbp]
\caption{LSTM classifier performance metrics.}
\begin{center}
\begin{tabular}{|c|c|c|c|}
\hline
\textbf{Precision}&\textbf{Recall} &\textbf{F1-Score}&\textbf{Accuracy}\\
\hline
1 & 0.35655 & 0.52568 & 0.35655 \\
\hline
\end{tabular}
\label{tab.lstm}
\end{center}
\end{table}

Due to imbalanced classes, we selected the f1-Score as the primary metric. Based on the results, the K-Nearest Neighbour classifier achieved the highest f1-Score at approximately 76\%, followed by Decision Trees and LSTM.

\section{Discussion}
\textcolor{black}{Through an analysis of historical data, our study identified three key factors (volume over 24 hours, maximum supply, and total supply) that could potentially influence cryptocurrency prices. However, it is essential to emphasize that the observed findings demonstrate correlation rather than causation between these identified factors and price fluctuations.}

\textcolor{black}{While our research did not specifically focus on market pump and dump schemes, a noteworthy observation is the potential association of a high deviation in the volume over 24 hours with such schemes. Our study revealed a positive correlation between the standard deviation of the price and both the mean and standard deviation of volume over 24 hours. Consequently, a higher standard deviation in the volume over 24 hours may serve as a potential indicator of elevated pump-and-dump activities.}

\textcolor{black}{Controlling the volume of a cryptocurrency to manipulate its price poses significant challenges, particularly when the tokens have high value and are widely distributed among users. However, the team that creates the cryptocurrency can exercise more feasible control over the other two parameters, namely circulating supply and maximum supply. In the absence of a maximum supply constraint, the team holds the potential to create an unlimited number of tokens, leading to inflation and subsequent depreciation of the token's value. Notably, the parameter of utmost importance is circulating supply, which may present challenges in its interpretation. For instance, a circulating supply value of 19 million may lack contextual significance by itself. Nevertheless, if this value represents 19 million out of a total supply of 20 million (assuming the maximum supply is defined and equal to the total supply), it indicates that the team retains control over only 1 million tokens. Conversely, if the total supply is, for instance, 100 million, the team has 81 million tokens yet to produce. Even if they were to produce only a portion, say 19 million tokens, it would result in a significant decrease in the token's value since it represents 100\% increase in the supply. }

\textcolor{black}{In order to be sure of the cryptocurrency team’s ability to manipulate the cryptocurrency when the circulating supply represent small portion of the total supply, we need to understand the production process of that cryptocurrency. Understanding the production process of tokens requires a thorough examination of the specific cryptocurrency's white paper and a manual review of its underlying code. However, it is essential to recognize that this process can be complicated, requiring the expertise of skilled professionals, and may not be feasible for all existing cryptocurrencies. As an alternative, evaluating the percentage of total supply in circulation emerges as a more practical approach. A higher percentage indicates that the cryptocurrency team's ability to manipulate token production is limited, while a lower percentage implies a greater potential for their influence on the market.}

\textcolor{black}{The classification of a cryptocurrency as "risky" through our classifiers does not definitively imply its certain demise in the future. Nonetheless, it serves as a warning signal, prompting investors to exercise increased caution and conduct thorough investigations before making investment decisions concerning such cryptocurrencies. This proactive approach to risk assessment can aid in making informed choices and contribute to a more stable and secure cryptocurrency investment landscape.
}

\section{Conclusion}
After analyzing historical cryptocurrency data from 2013 to January 1st, 2022, we discovered that nearly 39.34\% of cryptocurrencies vanished. Among the disappeared cryptocurrencies, 40\% had a lifetime of less than 80 days, and about 75\% of them had a lifetime of less than a year. We also observed that only approximately 10\% of existing cryptocurrencies have been in the market for over 1000 days.

In terms of measuring the correlation between cryptocurrency prices and other parameters, we found no linear correlation in the data between January 1st, 2020, and January 1st, 2022. \textcolor{black}{However, we did discover a strong negative Spearman correlation between maximum supply and price, a strong negative Spearman correlation between total supply and price, and a medium positive relationship between volume over 24 hours and price. By measuring the correlation between the standard deviation/mean of the price and other parameters' standard deviation/mean, we found a strong negative correlation with the mean of maximum supply and the mean of total supply. We also found that the Spearman correlation between the standard deviation/mean of the price and the standard deviation/mean of volume over 24 hours had a near-medium value.}

We utilized the K-means algorithm and the Elbow method to cluster cryptocurrencies based on their on-chain features. The optimal number of clusters was determined to be 5 using the Within-Cluster Sum of Square (WCSS) value. The PHP-ML library was used to normalize the features by dividing each one by its maximum value. The clustering process provides a better understanding of cryptocurrencies by comparing them with others in the same cluster.

Finally, we attempted to classify cryptocurrencies as either risky or not by training various classifiers on a historical dataset, with a disappeared cryptocurrency being classified as risky and an existing cryptocurrency as not risky. The K-Nearest Neighbor classifier yielded the best results with a score of approximately \textcolor{black}{76\%}. Additionally, we explored an LSTM architecture as it can treat the historical data of a cryptocurrency as a time series. However, the LSTM results were not satisfactory and require improvement in future work. We believe that using a more advanced model such as recurrent neural networks (RNN) \cite{medsker2001recurrent}, specifically, LSTM, is more appropriate and effective than traditional linear classifiers.

\section{Future Work}
Improving the classifiers used to detect risky cryptocurrencies can be achieved by addressing the issue of class imbalance. One way to improve the results is by using sampling techniques. Additionally, tuning the parameters of the classifiers can also lead to better results. Currently, the default settings provided by the library were used for each classifier \cite{scikit-learn}, except for the LSTM. Furthermore, it may be beneficial to use statistical data for each cryptocurrency instead of using all of their historical data, which could result in more accurate predictions.

\section*{Acknowledgment}
\noindent This work is supported partially by YTB (The Presidency of Turks Abroad and Related Communities) and TÜBİTAK (The Scientific and Technical Research Council of Türkiye) 2247-A National Leader Researchers Award 121C338. We would like to thank CoinMarketCap for providing us access to full history for special price.

\bibliographystyle{IEEEtran}
\bibliography{IEEEabrv,references}

\end{document}